\documentclass[%
 reprint,
 amsmath,amssymb,
 aps,
]{revtex4-2}

\usepackage{graphicx}
\usepackage{dcolumn}
\usepackage{bm}
\usepackage{float}
\usepackage{xcolor}
\usepackage{mhchem}
\usepackage{gensymb}
\usepackage{amsmath}

\begin{document}

\preprint{APS/123-QED}

\title{Constructing accurate machine-learned potentials and performing highly efficient atomistic simulations to predict structural and thermal properties}

\author{Junlan Liu\textsuperscript{1}}
\email{iris1025@seas.upenn.edu}
\affiliation{\textsuperscript{1}School of Engineering and Applied science, University of Pennsylvania, 220 South 33rd Street, Philadelphia, PA 19104-6391, USA}

\author{Qian Yin\textsuperscript{2}}
\affiliation{\textsuperscript{2}College of Resources and Environment, Chengdu University of Information Technology, Chengdu, Sichuan Province, 610225, China}

\author{Mengshu He\textsuperscript{4}}
\affiliation{\textsuperscript{4}USC Keck School of Medicine, University of Southern California, 1975 Zonal Ave, Los Angeles, CA, 90033, USA}
\email{carolinehe0029@gmail.com}

\author{Jun Zhou\textsuperscript{3}\textsuperscript{4}}
\email{zhouchaos@126.com}
\affiliation{\textsuperscript{3}College of Big Data and Intelligent Engineering, Southwest Forestry University, Kunming, Yunnan Province, 650224, China}
\affiliation{\textsuperscript{4}College
of Information Sciences and Technology, Donghua University, Shanghai 200051, China}

\date{\today}

\begin{abstract}
The $\text{Cu}_7\text{P}\text{S}_6$ compound has garnered significant attention due to its potential in thermoelectric applications. In this study, we introduce a neuroevolution potential (NEP), trained on a dataset generated from ab initio molecular dynamics (AIMD) simulations, using the moment tensor potential (MTP) as a reference. The low root mean square errors (RMSEs) for total energy and atomic forces demonstrate the high accuracy and transferability of both the MTP and NEP. We further calculate the phonon density of states (DOS) and radial distribution function (RDF) using both machine learning potentials, comparing the results to density functional theory (DFT) calculations. While the MTP potential offers slightly higher accuracy, the NEP achieves a remarkable 41-fold increase in computational speed. These findings provide detailed microscopic insights into the dynamics and rapid Cu-ion diffusion, paving the way for future studies on Cu-based solid electrolytes and their applications in energy devices. 
\end{abstract}

\maketitle

\section{INTRODUCTION}

Machine learning (ML) potentials represent a transformative breakthrough in computational materials science, offering an unprecedented combination of precision and efficiency for modeling complex atomic systems. These potentials harness the power of advanced machine learning algorithms to emulate the accuracy of \textit{ab initio} methods, such as density functional theory (DFT), while dramatically reducing computational costs. By accurately modeling interatomic interactions, ML potentials enable the exploration of large, intricate systems and long simulation timescales, surpassing the limitations of traditional quantum mechanical approaches.

The concept of ML potentials emerged in the early 1990s, with pioneering work fitting classical potentials using extensive \textit{ab initio} datasets~\cite{Ercolessi1992InteratomicPF, Ercolessi_1994}. Since then, significant advances have been achieved through the development of neural networks~\cite{Artrith2012HighdimensionalNN, Gastegger2015HighDimensionalNN, Dolgirev2016MachineLS, Smith2016ANI1AE, Zhang2017DeepPM, Pun2018PhysicallyIA} and kernel-based regression models~\cite{Schtt2013HowTR, Ramakrishnan2015ManyMP, Rupp2015MachineLF}, particularly Gaussian processes~\cite{Bartk2009GaussianAP, Deringer2016MachineLB, Grisafi2017SymmetryAdaptedML, Jinnouchi2019OntheflyML}. These approaches have yielded highly accurate potentials capable of predicting energies and forces with near \textit{ab initio} precision. Expansion-based models, such as the moment tensor potential (MTP)~\cite{Shapeev2015MomentTP} and other basis function-based approaches, represent another significant class of ML potentials, with notable implementations in tools like the MLIP package~\cite{Novikov2020TheMP, ouyang2022role}.

The versatility of ML potentials has unlocked new avenues for simulating key material properties. Applications include calculating mean squared displacements (MSD), phonon density of states (DOS), radial distribution functions (RDF), and lattice thermal conductivity through molecular dynamics (MD). For instance, Shen \textit{et al.} employed neural network potentials to investigate the lattice thermal conductivity of $\text{Cu}_7\text{P}\text{S}_6$~\cite{shen2024amorphous} and $\text{Ag}_8\text{Ge}\text{Se}_6$~\cite{shen2023soft}, 
while Wang \textit{et al.} explored the impact of fourth-order anharmonicity on the thermal properties of crystalline $ \rm CsPbBr_{3}$~\cite{wang2023role} and  Wang \textit{et al.} studied 2D phosphorous carbides \cite{cao2024thermal}
 using the homogeneous non-equilibrium molecular dynamics (HNEMD) method~\cite{fan2019homogeneous,evans1982homogeneous}.

In this study, we focus on $\text{Cu}_7\text{P}\text{S}_6$, a prominent superionic conductor (SIC) with high-temperature ionic conductivity, which holds significant potential for Cu-based battery applications. $\text{Cu}_7\text{P}\text{S}_6$ belongs to the argyrodite family of solid electrolytes~\cite{kuhs1979argyrodites}, adopting distinct structural phases at different temperatures. While the room-temperature cubic phase (P2\(_1\)3, No.198) exhibits high ionic conductivity, detailed studies on its structural and thermal properties remain limited~\cite{studenyak2017structural, beeken2008electrical}. Understanding these properties is critical for optimizing its application in energy storage technologies.

Molecular dynamics simulations are particularly well-suited for studying $\text{Cu}_7\text{P}\text{S}_6$, as they are not constrained by the order of anharmonic corrections. However, achieving accurate results requires precise interatomic potentials. Traditional force fields, such as reactive force field (ReaxFF)~\cite{senftle2016reaxff} or embedded atom methods~\cite{almyras2019semi}, often struggle to capture the complex behavior of SICs. In contrast, ML potentials have demonstrated their ability to achieve first-principles-level accuracy in a computationally efficient manner~\cite{butler2018machine, pun2019physically, qian2019thermal}.

In this work, we employ two advanced ML potentials---neuroevolution potential (NEP)~\cite{fan2021neuroevolution} and moment tensor potential (MTP)~\cite{Shapeev2015MomentTP}---to systematically investigate the structural properties of $\text{Cu}_7\text{P}\text{S}_6$. Key properties, such as DOS and RDF, are computed and compared with DFT benchmarks. While MTP demonstrates superior accuracy, NEP achieves computational speeds approximately 41 times faster, highlighting trade-offs between precision and efficiency. Our findings provide microscopic insights into the structural and thermal behavior of $\text{Cu}_7\text{P}\text{S}_6$, contributing to a deeper understanding of argyrodite-type superionic conductors and paving the way for their future applications.

\begin{figure*}
\centering
\includegraphics[width=1\linewidth]{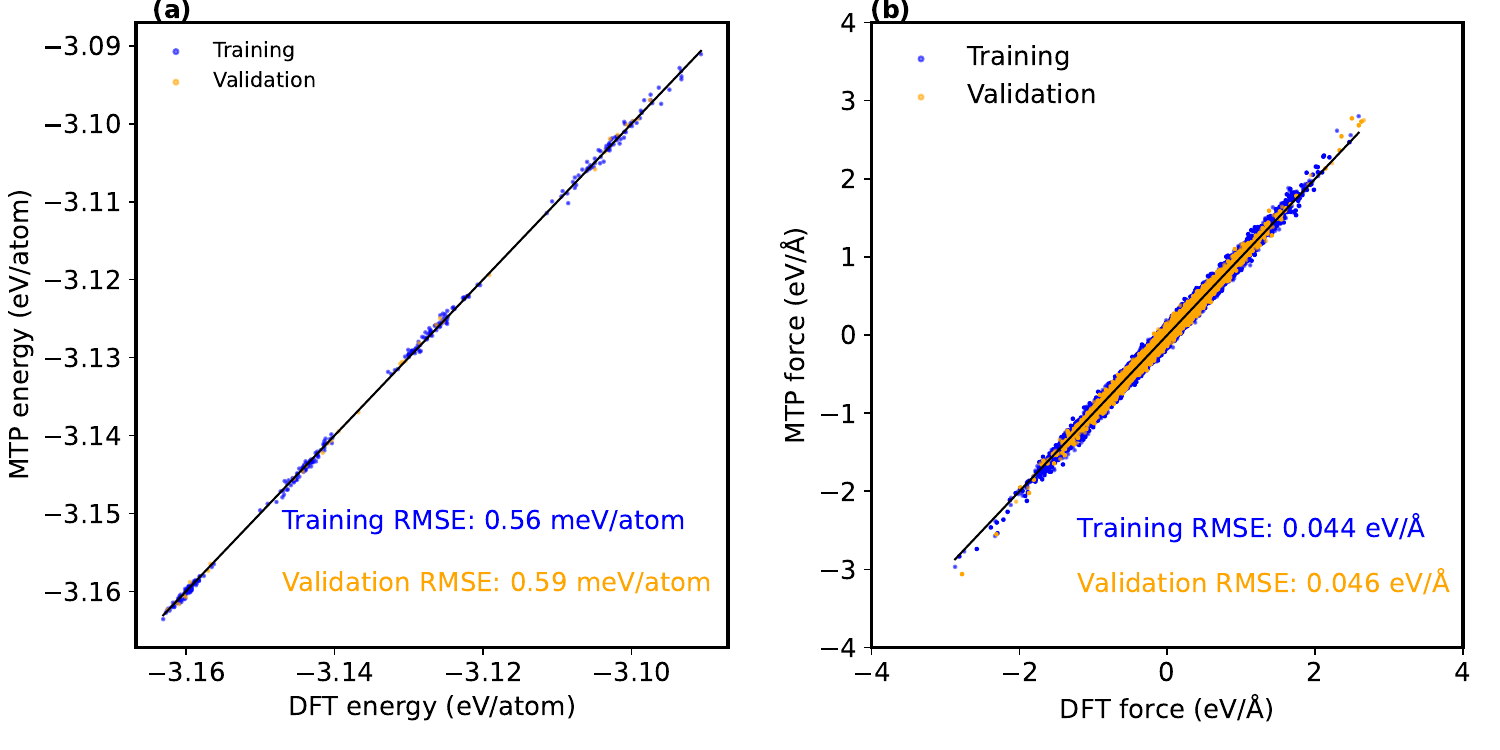}
\caption{\label{fig:wide} RMSEs of energies (a) and atomic forces (b) in the training and validation sets of $\text{Cu}_7\text{P}\text{S}_6$.}
\end{figure*}

\begin{figure*}
\centering
\includegraphics[width=1\linewidth]{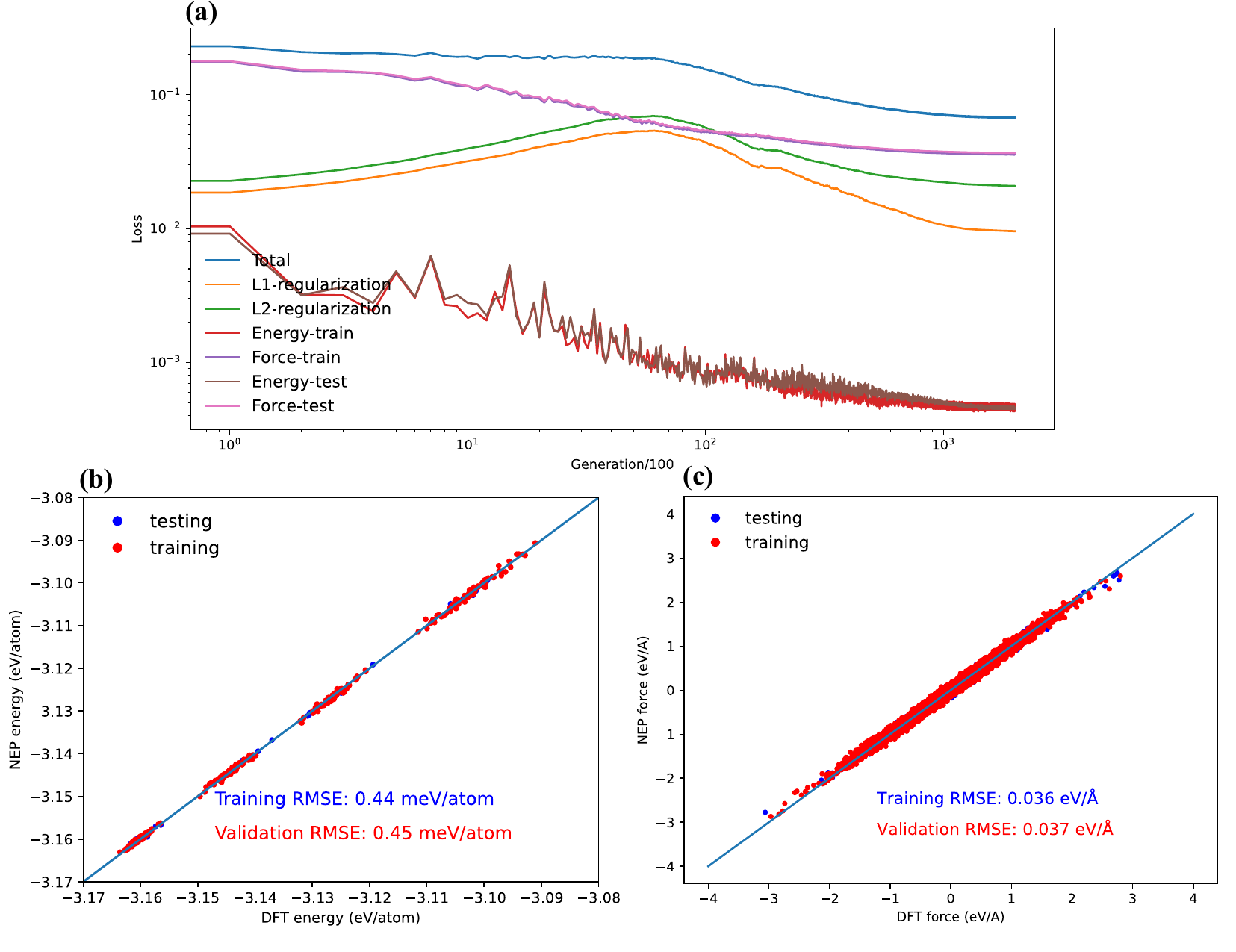}
\caption{\label{fig:wide} (a) The evolution of the loss functions (total, L1, L2, energy, and force) during the training process. (b) The energy calculated by NEP versus DFT. (c) The force calculated by NEP versus DFT.}
\end{figure*}

\begin{figure*}
\centering
\includegraphics[width=1\linewidth]{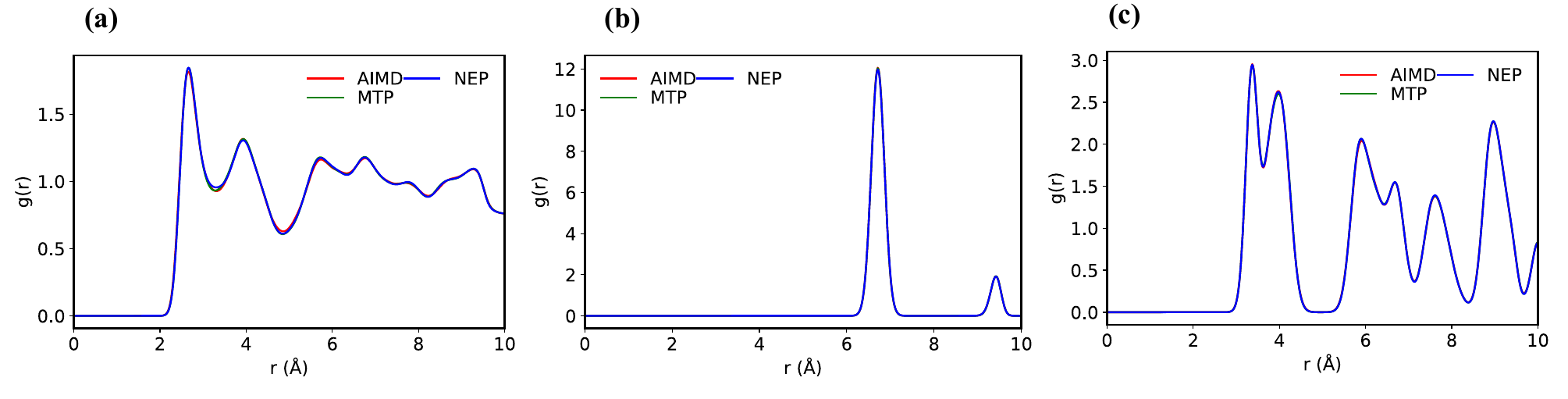}
\caption{\label{fig:wide} The calculated pair distribution function of different atomic pairs in $\text{Cu}_7\text{P}\text{S}_6$ at 300 K obtained from averaged over 200 ps MLMD trajectory with the MTP and NEP potentials, compared with AIMD simulations.}
\end{figure*}

\begin{figure*}
\centering
\includegraphics[width=1\linewidth]{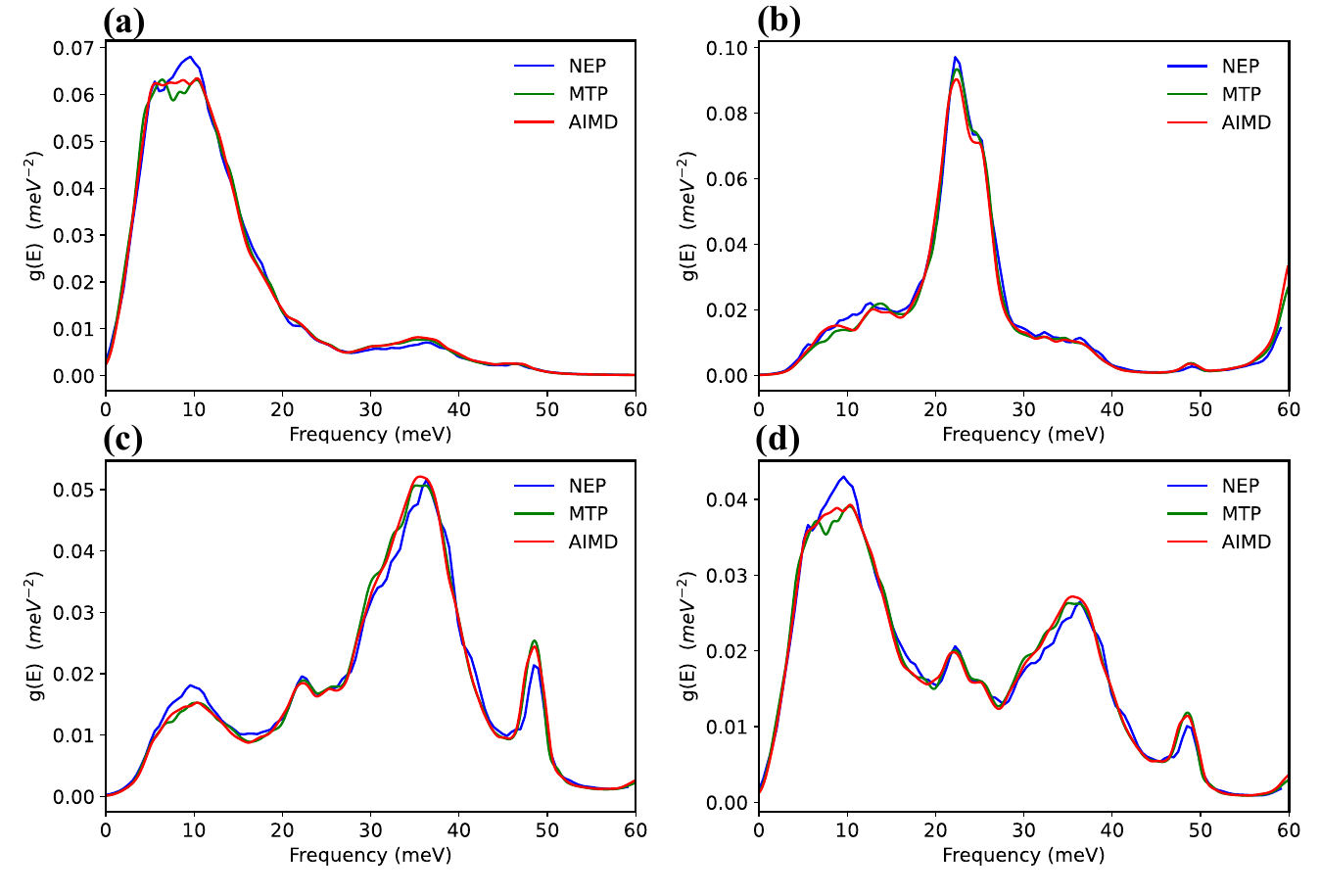}
\caption{\label{fig:wide} The calculated DOS of $\text{Cu}_7\text{P}\text{S}_6$ is obtained from the Fourier transform of the velocity autocorrelation function obtained from MLMD simulations with the MTP and NEP potentials at 300 K, and the results of AIMD calculations as a reference.}
\end{figure*}

\begin{figure}
\centering
\includegraphics[width=1\linewidth]{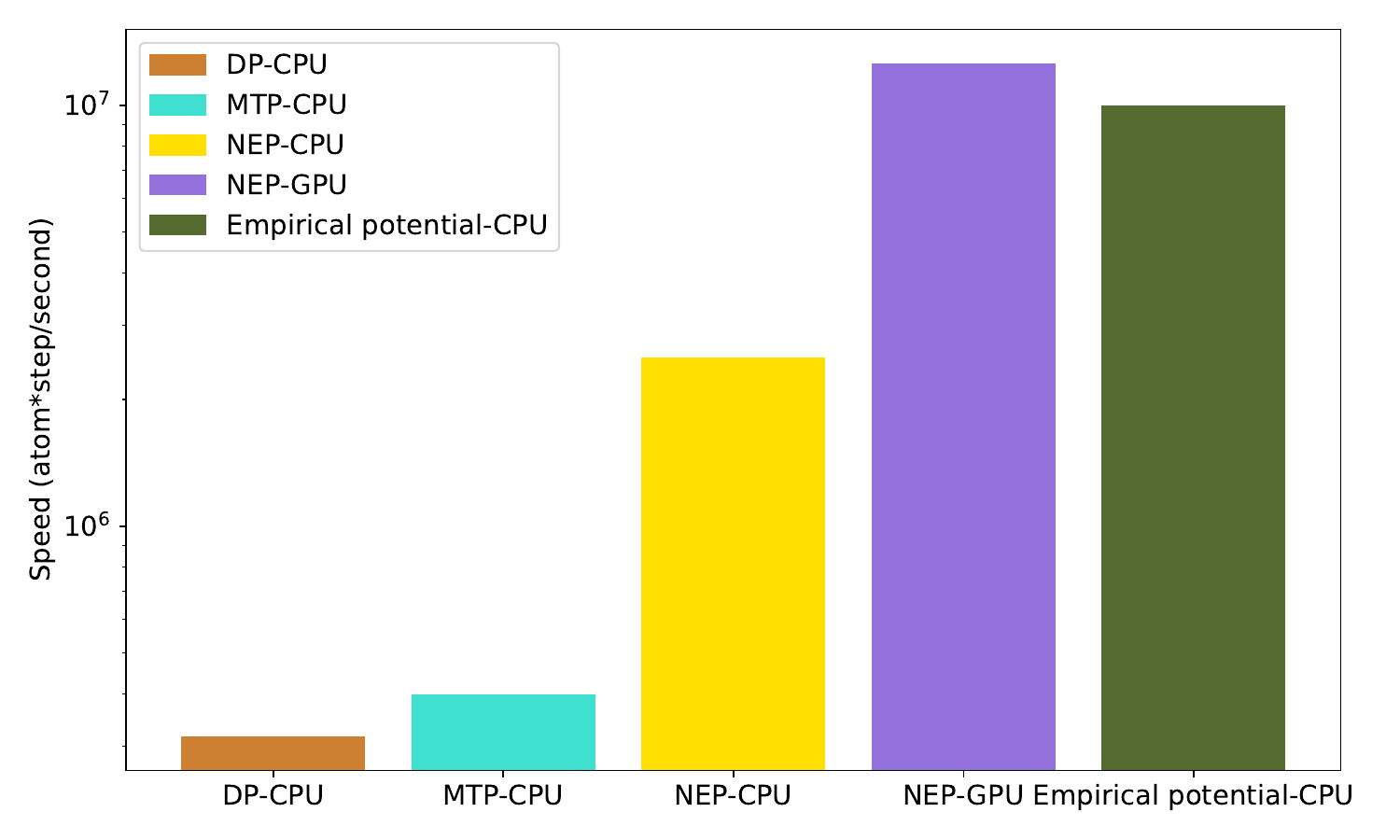}
\caption{\label{fig:wide} The computational speed of the NEP (running with one NVIDIA RTX 3080 implemented in GPUMD) is compared to those of the DP (running with one NVIDIA RTX 3080), NEP (CPU version), MTP, and Empirical potential (both running with 64 Intel Xeon Platinum 8375C CPU cores). Note that DP, MTP, and NEP (CPU version) are all implemented in the LAMMPS package\cite{plimpton1995fast}.}
\end{figure}

\section*{METHODS}

The electronic structure and energetics of $\text{Cu}_7\text{P}\text{S}_6$ were investigated using the Vienna \textit{ab initio} Simulation Package (VASP)~\cite{kresse1996efficient}, employing the projector augmented wave (PAW) method~\cite{blochl1994projector}. To ensure reliable results, the PBEsol exchange-correlation functional~\cite{perdew2008restoring} was chosen, as it provides lattice parameters in good agreement with experimental data~\cite{shen2023soft}. 

In self-consistent electronic calculations, a plane-wave energy cutoff of 500 eV was used alongside an energy convergence criterion of $10^{-8}$ eV. Calculations included the evaluation of Born effective charges and dielectric constants, which are crucial for characterizing the dynamical properties and ionic conductivity of the material.

The moment tensor potential (MTP)~\cite{Shapeev2015MomentTP} was constructed based on reference data generated through high-accuracy DFT calculations. To produce these configurations, ab initio molecular dynamics (AIMD) simulations were conducted using a 448-atom supercell across a temperature range up to 700 K under the NVT ensemble. The AIMD simulations employed a time step of 1 fs, with 1000 steps performed at each temperature. Structures for initial training were sampled every ten steps, yielding 300 configurations.

To further enrich the training dataset, active learning was applied in conjunction with MD simulations in the \textsc{LAMMPS} package~\cite{plimpton1995fast}. Simulations were performed at 300, 500, and 700 K using a time step of 1 fs. An extrapolation grade ($\gamma$) was used to monitor the representativeness of the configurations, with thresholds set at $\gamma_{\text{select}} = 3$ and $\gamma_{\text{break}} = 10$. Active learning simulations spanned 1 ns, an appropriate timescale for studying thermal properties~\cite{gubaev2019accelerating}. After active learning, a total of 354 representative configurations were selected for final MTP construction.

For these configurations, high-precision DFT calculations were performed using a $3 \times 3 \times 3 \, k$-point mesh with a total energy tolerance of $10^{-8}$ eV. These calculations yielded accurate energies, forces, and stresses to parametrize the MTP. The potential employed a cutoff radius of 6 Å to account for interatomic interactions.

The neuroevolution potential (NEP)~\cite{fan2021neuroevolution} was employed as an alternative approach to model atomic interactions with high accuracy and low computational cost. Radial and angular cutoffs were set to 7.0 Å and 5.0 Å, respectively, to define the interaction range. Training weights for atomic energies, forces, and stresses were set to $\omega_{e} = 1$, $\omega_{f} = 0.01$, and $\omega_{s} = 0.001$, respectively. To evaluate model performance, 10\% of the configurations were excluded from training and used as a validation set to ensure accuracy and transferability.

Phonon density of states (DOS) and radial distribution functions (RDF) were computed using molecular dynamics (MD) simulations in the GPUMD framework~\cite{dongre2017comparison, fan2019homogeneous}. These simulations employed a 56000-atom supercell to capture the material's superionic behavior accurately. The workflow included an equilibration phase followed by a 100 ps production run in the microcanonical (NVE) ensemble, with a correlation time of 20 ps. Before the simulations, energy minimization was performed to ensure stability.

NEP- and MTP-based MD simulations began with random-velocity initialization according to a Gaussian distribution, consistent with the target temperature. Time-averaging over sufficiently long trajectories ensured convergence of both temperature and energy to steady-state values. A time step of 1 fs was maintained, and periodic boundary conditions were applied in all three spatial directions. Multiple simulations were performed to minimize statistical uncertainties, providing robust phonon DOS and RDF results.

To ensure accuracy and reliability, all MD simulations underwent extensive convergence testing. The chosen potentials were benchmarked against DFT results for key thermodynamic and structural properties. Repeated simulations at varying initial conditions minimized bias and ensured reproducibility. The results highlight the balance between computational efficiency (NEP) and precision (MTP) \cite{ouyang2022temperature}, facilitating detailed insights into the thermal and structural behavior of $\text{Cu}_7\text{P}\text{S}_6$.

\section{RESULTS AND DISCUSSION}

\subsection*{Validation of Machine Learning Potentials}
To validate the performance of the trained machine learning potentials (MTP and NEP), 10\% of the configurations not included in the training set were reserved as the validation set. For both training and validation datasets, DFT calculations were conducted to evaluate the accuracy and transferability of the potentials. 

\textbf{Figure 1(a)} demonstrates that the predicted potential energies of the MTP align closely with DFT values for both training and validation sets, with root-mean-square errors (RMSEs) of 0.56 and 0.59 meV/atom, respectively. The distribution of residual errors (\textbf{Fig. 1(c)}) exhibits a zero-centered Gaussian profile, with more than 95\% of the validation configurations falling within $\pm$1.14 meV/atom. The consistency of atomic force predictions (\textbf{Fig. 1(b)}) further underscores the accuracy of the MTP, as indicated by RMSEs of 0.044 eV/Å (training) and 0.046 eV/Å (validation). The absence of error spikes even in high-energy configurations reflects the robustness of MTP for complex structures.

The NEP, evaluated through a similar protocol, shows even lower RMSEs for energy and force predictions. As illustrated in \textbf{Figures 2(b)} and \textbf{2(c)}, NEP achieves RMSEs of 0.44 and 0.45 meV/atom for training and validation sets, respectively, while maintaining RMSEs for atomic forces at 0.036 eV/Å (training) and 0.037 eV/Å (validation). The effective mitigation of overfitting during training (\textbf{Fig. 2(a)}) highlights the NEP's stability and transferability, positioning it as a highly reliable potential for $\text{Cu}_7\text{P}\text{S}_6$. Compared to MTP, NEP demonstrates superior force accuracy, making it particularly advantageous for simulations requiring precise interatomic interactions.

\subsection*{Structural Validation Using Radial Distribution Functions}
To assess the ability of MTP and NEP to replicate structural characteristics, radial distribution functions (RDFs) were computed from large-scale molecular dynamics (MD) simulations and compared with *ab initio* MD (AIMD) results. RDFs were obtained for both total and partial pair correlations at 300 K using a 448-atom supercell. 

As shown in \textbf{Figure 3}, RDFs predicted by NEP closely match those from AIMD simulations across all species, including Cu-Cu, Cu-S, and Cu-P interactions. Notably, NEP captures the subtle structural details with high fidelity, such as the positions and intensities of the first and second coordination peaks, which are crucial for describing local atomic environments in superionic conductors. While MTP predictions also align with AIMD, a slight underestimation of the Cu-Cu peak intensity is observed, indicating minor deviations in capturing short-range interactions. The consistency of NEP predictions across elemental contributions underscores its suitability for large-scale simulations of disordered systems like $\text{Cu}_7\text{P}\text{S}_6$.

\subsection*{Vibrational Properties from Phonon Density of States}
The phonon density of states (DOS), derived from the Fourier transform of velocity autocorrelation functions, was calculated at 300 K to analyze the vibrational spectrum of $\text{Cu}_7\text{P}\text{S}_6$. Comparisons among NEP, MTP, and AIMD-derived DOS are presented in \textbf{Figure 4}. The DOS curves computed with NEP and MTP show remarkable agreement with AIMD results, confirming the potentials' ability to capture vibrational dynamics accurately.

Low-energy modes, dominated by Cu vibrations, contribute to overlapping optical phonons below 10 meV, whereas intermediate and high-energy modes arise from P and S vibrations. NEP accurately reproduces these distinct phonon contributions, reflecting its robust interpolation of the potential energy surface. These results affirm the potential of NEP and MTP to describe both local and global vibrational properties, which are critical for evaluating thermal transport in superionic conductors.

\subsection*{Efficiency and Scalability of Machine Learning Potentials}
Beyond accuracy, computational efficiency is a key advantage of machine learning potentials. To quantify this, we compared the performance of NEP, MTP, and DeepMD potentials using MD simulations. \textbf{Figure 5} illustrates the execution times for each potential. NEP, implemented via the GPUMD package on an NVIDIA RTX 3080 GPU, is approximately 15 times faster than the GPU-accelerated DeepMD and orders of magnitude faster than CPU-only MTP simulations. This speed advantage, combined with high accuracy, positions NEP as a highly scalable tool for large-scale simulations.

The scalability of NEP, combined with its precision in energy and force predictions, enables simulations with tens of thousands of atoms, a scale previously infeasible with AIMD or even some other MLPs. This capability is particularly beneficial for studying temperature-dependent dynamics and long-wavelength phonon transport, critical for designing advanced ionic conductors.

\subsection*{Implications for Superionic Conductors}
The ability of NEP and MTP to reproduce structural and vibrational properties with high accuracy has broader implications for understanding superionic conductors. Accurate modeling of RDFs and DOS allows for insights into ion migration mechanisms and thermal transport phenomena. For instance, the vibrational modes captured by NEP directly inform the coupling between Cu ions and the lattice framework, a critical factor in superionic conduction. Furthermore, NEP’s efficiency enables the simulation of dynamic processes over extended timescales, offering a pathway for studying emergent phenomena such as defect migration and anharmonic effects in $\text{Cu}_7\text{P}\text{S}_6$ and related materials.

\section{CONCLUSION}
In this work, we developed and benchmarked NEP and MTP potentials using a comprehensive dataset derived from AIMD simulations to investigate the structural and thermal properties of the argyrodite-type superionic conductor $\text{Cu}_7\text{P}\text{S}_6$. The exceptionally low RMSEs for total energy and atomic forces confirm the high accuracy and predictive power of both machine learning potentials across diverse atomic configurations. We then employed both potentials to compute phonon DOS and RDFs, which were systematically compared to reference DFT calculations. While MTP demonstrated marginally higher accuracy in reproducing fine structural details, NEP achieved comparable fidelity with a significant computational advantage—operating approximately 41 times faster than MTP. This efficiency positions NEP as an ideal tool for large-scale and long-timescale simulations of superionic conductors, where the dynamics of fast ion diffusion are critical.
These findings reveal the capability of NEP and MTP to capture the essential microscopic mechanisms underlying rapid Cu-ion migration and vibrational behaviors in $\text{Cu}_7\text{P}\text{S}_6$. By providing accurate and scalable tools for modeling such systems, this study paves the way for exploring a broader class of argyrodite-type conductors and optimizing their properties for next-generation energy storage and conversion applications.

\begin{acknowledgments}
This work is partially supported by Yunnan Fundamental Research Projects (No.202001AT070112).
\end{acknowledgments}

\bibliography{apssamp}

\providecommand{\noopsort}[1]{}\providecommand{\singleletter}[1]{#1}%
\begin{thebibliography}{40}%
\makeatletter
\providecommand \@ifxundefined [1]{%
 \@ifx{#1\undefined}
}%
\providecommand \@ifnum [1]{%
 \ifnum #1\expandafter \@firstoftwo
 \else \expandafter \@secondoftwo
 \fi
}%
\providecommand \@ifx [1]{%
 \ifx #1\expandafter \@firstoftwo
 \else \expandafter \@secondoftwo
 \fi
}%
\providecommand \natexlab [1]{#1}%
\providecommand \enquote  [1]{``#1''}%
\providecommand \bibnamefont  [1]{#1}%
\providecommand \bibfnamefont [1]{#1}%
\providecommand \citenamefont [1]{#1}%
\providecommand \href@noop [0]{\@secondoftwo}%
\providecommand \href [0]{\begingroup \@sanitize@url \@href}%
\providecommand \@href[1]{\@@startlink{#1}\@@href}%
\providecommand \@@href[1]{\endgroup#1\@@endlink}%
\providecommand \@sanitize@url [0]{\catcode `\\12\catcode `\$12\catcode `\&12\catcode `\#12\catcode `\^12\catcode `\_12\catcode `\%12\relax}%
\providecommand \@@startlink[1]{}%
\providecommand \@@endlink[0]{}%
\providecommand \url  [0]{\begingroup\@sanitize@url \@url }%
\providecommand \@url [1]{\endgroup\@href {#1}{\urlprefix }}%
\providecommand \urlprefix  [0]{URL }%
\providecommand \Eprint [0]{\href }%
\providecommand \doibase [0]{https://doi.org/}%
\providecommand \selectlanguage [0]{\@gobble}%
\providecommand \bibinfo  [0]{\@secondoftwo}%
\providecommand \bibfield  [0]{\@secondoftwo}%
\providecommand \translation [1]{[#1]}%
\providecommand \BibitemOpen [0]{}%
\providecommand \bibitemStop [0]{}%
\providecommand \bibitemNoStop [0]{.\EOS\space}%
\providecommand \EOS [0]{\spacefactor3000\relax}%
\providecommand \BibitemShut  [1]{\csname bibitem#1\endcsname}%
\let\auto@bib@innerbib\@empty
\bibitem [{\citenamefont {Ercolessi}\ and\ \citenamefont {Adams}(1992)}]{Ercolessi1992InteratomicPF}%
  \BibitemOpen
  \bibfield  {author} {\bibinfo {author} {\bibfnamefont {F.}~\bibnamefont {Ercolessi}}\ and\ \bibinfo {author} {\bibfnamefont {J.~B.}\ \bibnamefont {Adams}},\ }\bibfield  {title} {\bibinfo {title} {Interatomic potentials from first-principles calculations},\ }\href {https://api.semanticscholar.org/CorpusID:95391806} {\bibfield  {journal} {\bibinfo  {journal} {MRS Proceedings}\ }\textbf {\bibinfo {volume} {291}},\ \bibinfo {pages} {31} (\bibinfo {year} {1992})}\BibitemShut {NoStop}%
\bibitem [{\citenamefont {Ercolessi}\ and\ \citenamefont {Adams}(1994)}]{Ercolessi_1994}%
  \BibitemOpen
  \bibfield  {author} {\bibinfo {author} {\bibfnamefont {F.}~\bibnamefont {Ercolessi}}\ and\ \bibinfo {author} {\bibfnamefont {J.~B.}\ \bibnamefont {Adams}},\ }\bibfield  {title} {\bibinfo {title} {Interatomic potentials from first-principles calculations: The force-matching method},\ }\href {https://doi.org/10.1209/0295-5075/26/8/005} {\bibfield  {journal} {\bibinfo  {journal} {Europhysics Letters}\ }\textbf {\bibinfo {volume} {26}},\ \bibinfo {pages} {583} (\bibinfo {year} {1994})}\BibitemShut {NoStop}%
\bibitem [{\citenamefont {Artrith}\ and\ \citenamefont {Behler}(2012)}]{Artrith2012HighdimensionalNN}%
  \BibitemOpen
  \bibfield  {author} {\bibinfo {author} {\bibfnamefont {N.}~\bibnamefont {Artrith}}\ and\ \bibinfo {author} {\bibfnamefont {J.}~\bibnamefont {Behler}},\ }\bibfield  {title} {\bibinfo {title} {High-dimensional neural network potentials for metal surfaces: A prototype study for copper},\ }\href {https://api.semanticscholar.org/CorpusID:54058794} {\bibfield  {journal} {\bibinfo  {journal} {Physical Review B}\ }\textbf {\bibinfo {volume} {85}},\ \bibinfo {pages} {045439} (\bibinfo {year} {2012})}\BibitemShut {NoStop}%
\bibitem [{\citenamefont {Gastegger}\ and\ \citenamefont {Marquetand}(2015)}]{Gastegger2015HighDimensionalNN}%
  \BibitemOpen
  \bibfield  {author} {\bibinfo {author} {\bibfnamefont {M.}~\bibnamefont {Gastegger}}\ and\ \bibinfo {author} {\bibfnamefont {P.}~\bibnamefont {Marquetand}},\ }\bibfield  {title} {\bibinfo {title} {High-dimensional neural network potentials for organic reactions and an improved training algorithm.},\ }\href {https://api.semanticscholar.org/CorpusID:21150039} {\bibfield  {journal} {\bibinfo  {journal} {Journal of chemical theory and computation}\ }\textbf {\bibinfo {volume} {11 5}},\ \bibinfo {pages} {2187} (\bibinfo {year} {2015})}\BibitemShut {NoStop}%
\bibitem [{\citenamefont {Dolgirev}\ \emph {et~al.}(2016)\citenamefont {Dolgirev}, \citenamefont {Kruglov},\ and\ \citenamefont {Oganov}}]{Dolgirev2016MachineLS}%
  \BibitemOpen
  \bibfield  {author} {\bibinfo {author} {\bibfnamefont {P.~E.}\ \bibnamefont {Dolgirev}}, \bibinfo {author} {\bibfnamefont {I.~A.}\ \bibnamefont {Kruglov}},\ and\ \bibinfo {author} {\bibfnamefont {A.~R.}\ \bibnamefont {Oganov}},\ }\bibfield  {title} {\bibinfo {title} {Machine learning scheme for fast extraction of chemically interpretable interatomic potentials},\ }\href {https://api.semanticscholar.org/CorpusID:35100520} {\bibfield  {journal} {\bibinfo  {journal} {AIP Advances}\ }\textbf {\bibinfo {volume} {6}},\ \bibinfo {pages} {085318} (\bibinfo {year} {2016})}\BibitemShut {NoStop}%
\bibitem [{\citenamefont {Smith}\ \emph {et~al.}(2016)\citenamefont {Smith}, \citenamefont {Isayev},\ and\ \citenamefont {Roitberg}}]{Smith2016ANI1AE}%
  \BibitemOpen
  \bibfield  {author} {\bibinfo {author} {\bibfnamefont {J.~S.}\ \bibnamefont {Smith}}, \bibinfo {author} {\bibfnamefont {O.}~\bibnamefont {Isayev}},\ and\ \bibinfo {author} {\bibfnamefont {A.~E.}\ \bibnamefont {Roitberg}},\ }\bibfield  {title} {\bibinfo {title} {Ani-1: an extensible neural network potential with dft accuracy at force field computational cost},\ }\href {https://api.semanticscholar.org/CorpusID:24604537} {\bibfield  {journal} {\bibinfo  {journal} {Chemical Science}\ }\textbf {\bibinfo {volume} {8}},\ \bibinfo {pages} {3192 } (\bibinfo {year} {2016})}\BibitemShut {NoStop}%
\bibitem [{\citenamefont {Zhang}\ \emph {et~al.}(2017)\citenamefont {Zhang}, \citenamefont {Han}, \citenamefont {Wang}, \citenamefont {Car},\ and\ \citenamefont {Weinan}}]{Zhang2017DeepPM}%
  \BibitemOpen
  \bibfield  {author} {\bibinfo {author} {\bibfnamefont {L.}~\bibnamefont {Zhang}}, \bibinfo {author} {\bibfnamefont {J.}~\bibnamefont {Han}}, \bibinfo {author} {\bibfnamefont {H.}~\bibnamefont {Wang}}, \bibinfo {author} {\bibfnamefont {R.}~\bibnamefont {Car}},\ and\ \bibinfo {author} {\bibfnamefont {E.}~\bibnamefont {Weinan}},\ }\bibfield  {title} {\bibinfo {title} {Deep potential molecular dynamics: a scalable model with the accuracy of quantum mechanics},\ }\href {https://api.semanticscholar.org/CorpusID:19098240} {\bibfield  {journal} {\bibinfo  {journal} {Physical review letters}\ }\textbf {\bibinfo {volume} {120 14}},\ \bibinfo {pages} {143001} (\bibinfo {year} {2017})}\BibitemShut {NoStop}%
\bibitem [{\citenamefont {Pun}\ \emph {et~al.}(2018)\citenamefont {Pun}, \citenamefont {Batra}, \citenamefont {Ramprasad},\ and\ \citenamefont {Mishin}}]{Pun2018PhysicallyIA}%
  \BibitemOpen
  \bibfield  {author} {\bibinfo {author} {\bibfnamefont {G.~P.~P.}\ \bibnamefont {Pun}}, \bibinfo {author} {\bibfnamefont {R.}~\bibnamefont {Batra}}, \bibinfo {author} {\bibfnamefont {R.}~\bibnamefont {Ramprasad}},\ and\ \bibinfo {author} {\bibfnamefont {Y.}~\bibnamefont {Mishin}},\ }\bibfield  {title} {\bibinfo {title} {Physically informed artificial neural networks for atomistic modeling of materials},\ }\href {https://api.semanticscholar.org/CorpusID:119071683} {\bibfield  {journal} {\bibinfo  {journal} {Nature Communications}\ }\textbf {\bibinfo {volume} {10}} (\bibinfo {year} {2018})}\BibitemShut {NoStop}%
\bibitem [{\citenamefont {Sch{\"u}tt}\ \emph {et~al.}(2013)\citenamefont {Sch{\"u}tt}, \citenamefont {Glawe}, \citenamefont {Brockherde}, \citenamefont {Brockherde}, \citenamefont {Sanna}, \citenamefont {Muller}, \citenamefont {Muller},\ and\ \citenamefont {Gross}}]{Schtt2013HowTR}%
  \BibitemOpen
  \bibfield  {author} {\bibinfo {author} {\bibfnamefont {K.~T.}\ \bibnamefont {Sch{\"u}tt}}, \bibinfo {author} {\bibfnamefont {H.}~\bibnamefont {Glawe}}, \bibinfo {author} {\bibfnamefont {F.}~\bibnamefont {Brockherde}}, \bibinfo {author} {\bibfnamefont {F.}~\bibnamefont {Brockherde}}, \bibinfo {author} {\bibfnamefont {A.}~\bibnamefont {Sanna}}, \bibinfo {author} {\bibfnamefont {K.}~\bibnamefont {Muller}}, \bibinfo {author} {\bibfnamefont {K.}~\bibnamefont {Muller}},\ and\ \bibinfo {author} {\bibfnamefont {E.~K.~U.}\ \bibnamefont {Gross}},\ }\bibfield  {title} {\bibinfo {title} {How to represent crystal structures for machine learning: Towards fast prediction of electronic properties},\ }\href {https://api.semanticscholar.org/CorpusID:44054738} {\bibfield  {journal} {\bibinfo  {journal} {Physical Review B}\ }\textbf {\bibinfo {volume} {89}},\ \bibinfo {pages} {205118} (\bibinfo {year} {2013})}\BibitemShut {NoStop}%
\bibitem [{\citenamefont {Ramakrishnan}\ and\ \citenamefont {von Lilienfeld}(2015)}]{Ramakrishnan2015ManyMP}%
  \BibitemOpen
  \bibfield  {author} {\bibinfo {author} {\bibfnamefont {R.}~\bibnamefont {Ramakrishnan}}\ and\ \bibinfo {author} {\bibfnamefont {O.~A.}\ \bibnamefont {von Lilienfeld}},\ }\bibfield  {title} {\bibinfo {title} {Many molecular properties from one kernel in chemical space.},\ }\href {https://api.semanticscholar.org/CorpusID:13996962} {\bibfield  {journal} {\bibinfo  {journal} {Chimia}\ }\textbf {\bibinfo {volume} {69 4}},\ \bibinfo {pages} {182} (\bibinfo {year} {2015})}\BibitemShut {NoStop}%
\bibitem [{\citenamefont {Rupp}(2015)}]{Rupp2015MachineLF}%
  \BibitemOpen
  \bibfield  {author} {\bibinfo {author} {\bibfnamefont {M.}~\bibnamefont {Rupp}},\ }\bibfield  {title} {\bibinfo {title} {Machine learning for quantum mechanics in a nutshell},\ }\href {https://api.semanticscholar.org/CorpusID:94314605} {\bibfield  {journal} {\bibinfo  {journal} {International Journal of Quantum Chemistry}\ }\textbf {\bibinfo {volume} {115}},\ \bibinfo {pages} {1058} (\bibinfo {year} {2015})}\BibitemShut {NoStop}%
\bibitem [{\citenamefont {Bart{\'o}k}\ \emph {et~al.}(2009)\citenamefont {Bart{\'o}k}, \citenamefont {Payne}, \citenamefont {Kondor},\ and\ \citenamefont {Cs{\'a}nyi}}]{Bartk2009GaussianAP}%
  \BibitemOpen
  \bibfield  {author} {\bibinfo {author} {\bibfnamefont {A.~P.}\ \bibnamefont {Bart{\'o}k}}, \bibinfo {author} {\bibfnamefont {M.~C.}\ \bibnamefont {Payne}}, \bibinfo {author} {\bibfnamefont {R.}~\bibnamefont {Kondor}},\ and\ \bibinfo {author} {\bibfnamefont {G.}~\bibnamefont {Cs{\'a}nyi}},\ }\bibfield  {title} {\bibinfo {title} {Gaussian approximation potentials: the accuracy of quantum mechanics, without the electrons.},\ }\href {https://api.semanticscholar.org/CorpusID:15918457} {\bibfield  {journal} {\bibinfo  {journal} {Physical review letters}\ }\textbf {\bibinfo {volume} {104 13}},\ \bibinfo {pages} {136403} (\bibinfo {year} {2009})}\BibitemShut {NoStop}%
\bibitem [{\citenamefont {Deringer}\ and\ \citenamefont {Cs{\'a}nyi}(2016)}]{Deringer2016MachineLB}%
  \BibitemOpen
  \bibfield  {author} {\bibinfo {author} {\bibfnamefont {V.~L.}\ \bibnamefont {Deringer}}\ and\ \bibinfo {author} {\bibfnamefont {G.}~\bibnamefont {Cs{\'a}nyi}},\ }\bibfield  {title} {\bibinfo {title} {Machine learning based interatomic potential for amorphous carbon},\ }\href {https://api.semanticscholar.org/CorpusID:55190594} {\bibfield  {journal} {\bibinfo  {journal} {Physical Review B}\ }\textbf {\bibinfo {volume} {95}},\ \bibinfo {pages} {094203} (\bibinfo {year} {2016})}\BibitemShut {NoStop}%
\bibitem [{\citenamefont {Grisafi}\ \emph {et~al.}(2017)\citenamefont {Grisafi}, \citenamefont {Wilkins}, \citenamefont {Cs{\'a}nyi},\ and\ \citenamefont {Ceriotti}}]{Grisafi2017SymmetryAdaptedML}%
  \BibitemOpen
  \bibfield  {author} {\bibinfo {author} {\bibfnamefont {A.}~\bibnamefont {Grisafi}}, \bibinfo {author} {\bibfnamefont {D.~M.}\ \bibnamefont {Wilkins}}, \bibinfo {author} {\bibfnamefont {G.}~\bibnamefont {Cs{\'a}nyi}},\ and\ \bibinfo {author} {\bibfnamefont {M.}~\bibnamefont {Ceriotti}},\ }\bibfield  {title} {\bibinfo {title} {Symmetry-adapted machine learning for tensorial properties of atomistic systems.},\ }\href {https://api.semanticscholar.org/CorpusID:46764595} {\bibfield  {journal} {\bibinfo  {journal} {Physical review letters}\ }\textbf {\bibinfo {volume} {120 3}},\ \bibinfo {pages} {036002} (\bibinfo {year} {2017})}\BibitemShut {NoStop}%
\bibitem [{\citenamefont {Jinnouchi}\ \emph {et~al.}(2019)\citenamefont {Jinnouchi}, \citenamefont {Karsai},\ and\ \citenamefont {Kresse}}]{Jinnouchi2019OntheflyML}%
  \BibitemOpen
  \bibfield  {author} {\bibinfo {author} {\bibfnamefont {R.}~\bibnamefont {Jinnouchi}}, \bibinfo {author} {\bibfnamefont {F.}~\bibnamefont {Karsai}},\ and\ \bibinfo {author} {\bibfnamefont {G.}~\bibnamefont {Kresse}},\ }\bibfield  {title} {\bibinfo {title} {On-the-fly machine learning force field generation: Application to melting points},\ }\href {https://api.semanticscholar.org/CorpusID:140220650} {\bibfield  {journal} {\bibinfo  {journal} {Physical Review B}\ } (\bibinfo {year} {2019})}\BibitemShut {NoStop}%
\bibitem [{\citenamefont {Shapeev}(2015)}]{Shapeev2015MomentTP}%
  \BibitemOpen
  \bibfield  {author} {\bibinfo {author} {\bibfnamefont {A.~V.}\ \bibnamefont {Shapeev}},\ }\bibfield  {title} {\bibinfo {title} {Moment tensor potentials: A class of systematically improvable interatomic potentials},\ }\href {https://api.semanticscholar.org/CorpusID:28970251} {\bibfield  {journal} {\bibinfo  {journal} {Multiscale Model. Simul.}\ }\textbf {\bibinfo {volume} {14}},\ \bibinfo {pages} {1153} (\bibinfo {year} {2015})}\BibitemShut {NoStop}%
\bibitem [{\citenamefont {Novikov}\ \emph {et~al.}(2020)\citenamefont {Novikov}, \citenamefont {Gubaev}, \citenamefont {Podryabinkin},\ and\ \citenamefont {Shapeev}}]{Novikov2020TheMP}%
  \BibitemOpen
  \bibfield  {author} {\bibinfo {author} {\bibfnamefont {I.~S.}\ \bibnamefont {Novikov}}, \bibinfo {author} {\bibfnamefont {K.}~\bibnamefont {Gubaev}}, \bibinfo {author} {\bibfnamefont {E.~V.}\ \bibnamefont {Podryabinkin}},\ and\ \bibinfo {author} {\bibfnamefont {A.~V.}\ \bibnamefont {Shapeev}},\ }\bibfield  {title} {\bibinfo {title} {The mlip package: moment tensor potentials with mpi and active learning},\ }\href {https://api.semanticscholar.org/CorpusID:220633602} {\bibfield  {journal} {\bibinfo  {journal} {Machine Learning: Science and Technology}\ }\textbf {\bibinfo {volume} {2}} (\bibinfo {year} {2020})}\BibitemShut {NoStop}%
\bibitem [{\citenamefont {Ouyang}\ \emph {et~al.}(2022{\natexlab{a}})\citenamefont {Ouyang}, \citenamefont {Wang},\ and\ \citenamefont {Chen}}]{ouyang2022role}%
  \BibitemOpen
  \bibfield  {author} {\bibinfo {author} {\bibfnamefont {N.}~\bibnamefont {Ouyang}}, \bibinfo {author} {\bibfnamefont {C.}~\bibnamefont {Wang}},\ and\ \bibinfo {author} {\bibfnamefont {Y.}~\bibnamefont {Chen}},\ }\bibfield  {title} {\bibinfo {title} {Role of alloying in the phonon and thermal transport of sns--snse across the phase transition},\ }\href@noop {} {\bibfield  {journal} {\bibinfo  {journal} {Materials Today Physics}\ }\textbf {\bibinfo {volume} {28}},\ \bibinfo {pages} {100890} (\bibinfo {year} {2022}{\natexlab{a}})}\BibitemShut {NoStop}%
\bibitem [{\citenamefont {Shen}\ \emph {et~al.}(2024)\citenamefont {Shen}, \citenamefont {Ouyang}, \citenamefont {Huang}, \citenamefont {Tung}, \citenamefont {Yang}, \citenamefont {Faizan}, \citenamefont {Perez}, \citenamefont {He}, \citenamefont {Sotnikov}, \citenamefont {Willa} \emph {et~al.}}]{shen2024amorphous}%
  \BibitemOpen
  \bibfield  {author} {\bibinfo {author} {\bibfnamefont {X.}~\bibnamefont {Shen}}, \bibinfo {author} {\bibfnamefont {N.}~\bibnamefont {Ouyang}}, \bibinfo {author} {\bibfnamefont {Y.}~\bibnamefont {Huang}}, \bibinfo {author} {\bibfnamefont {Y.-H.}\ \bibnamefont {Tung}}, \bibinfo {author} {\bibfnamefont {C.-C.}\ \bibnamefont {Yang}}, \bibinfo {author} {\bibfnamefont {M.}~\bibnamefont {Faizan}}, \bibinfo {author} {\bibfnamefont {N.}~\bibnamefont {Perez}}, \bibinfo {author} {\bibfnamefont {R.}~\bibnamefont {He}}, \bibinfo {author} {\bibfnamefont {A.}~\bibnamefont {Sotnikov}}, \bibinfo {author} {\bibfnamefont {K.}~\bibnamefont {Willa}}, \emph {et~al.},\ }\bibfield  {title} {\bibinfo {title} {Amorphous-like ultralow thermal transport in crystalline argyrodite cu7ps6},\ }\href@noop {} {\bibfield  {journal} {\bibinfo  {journal} {Advanced Science}\ ,\ \bibinfo {pages} {2400258}} (\bibinfo {year} {2024})}\BibitemShut {NoStop}%
\bibitem [{\citenamefont {Shen}\ \emph {et~al.}(2023)\citenamefont {Shen}, \citenamefont {Koza}, \citenamefont {Tung}, \citenamefont {Ouyang}, \citenamefont {Yang}, \citenamefont {Wang}, \citenamefont {Chen}, \citenamefont {Willa}, \citenamefont {Heid}, \citenamefont {Zhou} \emph {et~al.}}]{shen2023soft}%
  \BibitemOpen
  \bibfield  {author} {\bibinfo {author} {\bibfnamefont {X.}~\bibnamefont {Shen}}, \bibinfo {author} {\bibfnamefont {M.~M.}\ \bibnamefont {Koza}}, \bibinfo {author} {\bibfnamefont {Y.-H.}\ \bibnamefont {Tung}}, \bibinfo {author} {\bibfnamefont {N.}~\bibnamefont {Ouyang}}, \bibinfo {author} {\bibfnamefont {C.-C.}\ \bibnamefont {Yang}}, \bibinfo {author} {\bibfnamefont {C.}~\bibnamefont {Wang}}, \bibinfo {author} {\bibfnamefont {Y.}~\bibnamefont {Chen}}, \bibinfo {author} {\bibfnamefont {K.}~\bibnamefont {Willa}}, \bibinfo {author} {\bibfnamefont {R.}~\bibnamefont {Heid}}, \bibinfo {author} {\bibfnamefont {X.}~\bibnamefont {Zhou}}, \emph {et~al.},\ }\bibfield  {title} {\bibinfo {title} {Soft phonon mode triggering fast ag diffusion in superionic argyrodite ag8gese6},\ }\href@noop {} {\bibfield  {journal} {\bibinfo  {journal} {Small}\ }\textbf {\bibinfo {volume} {19}},\ \bibinfo {pages} {2305048} (\bibinfo {year} {2023})}\BibitemShut {NoStop}%
\bibitem [{\citenamefont {Wang}\ \emph {et~al.}(2023)\citenamefont {Wang}, \citenamefont {Gao}, \citenamefont {Zhu}, \citenamefont {Ren}, \citenamefont {Hu}, \citenamefont {Sun}, \citenamefont {Ding}, \citenamefont {Xia},\ and\ \citenamefont {Li}}]{wang2023role}%
  \BibitemOpen
  \bibfield  {author} {\bibinfo {author} {\bibfnamefont {X.}~\bibnamefont {Wang}}, \bibinfo {author} {\bibfnamefont {Z.}~\bibnamefont {Gao}}, \bibinfo {author} {\bibfnamefont {G.}~\bibnamefont {Zhu}}, \bibinfo {author} {\bibfnamefont {J.}~\bibnamefont {Ren}}, \bibinfo {author} {\bibfnamefont {L.}~\bibnamefont {Hu}}, \bibinfo {author} {\bibfnamefont {J.}~\bibnamefont {Sun}}, \bibinfo {author} {\bibfnamefont {X.}~\bibnamefont {Ding}}, \bibinfo {author} {\bibfnamefont {Y.}~\bibnamefont {Xia}},\ and\ \bibinfo {author} {\bibfnamefont {B.}~\bibnamefont {Li}},\ }\bibfield  {title} {\bibinfo {title} {Role of high-order anharmonicity and off-diagonal terms in thermal conductivity: A case study of multiphase cspbbr 3},\ }\href@noop {} {\bibfield  {journal} {\bibinfo  {journal} {Physical Review B}\ }\textbf {\bibinfo {volume} {107}},\ \bibinfo {pages} {214308} (\bibinfo {year} {2023})}\BibitemShut {NoStop}%
\bibitem [{\citenamefont {Cao}\ \emph {et~al.}(2024)\citenamefont {Cao}, \citenamefont {Cao}, \citenamefont {Zhu}, \citenamefont {Dong}, \citenamefont {Wang},\ and\ \citenamefont {Qian}}]{cao2024thermal}%
  \BibitemOpen
  \bibfield  {author} {\bibinfo {author} {\bibfnamefont {C.}~\bibnamefont {Cao}}, \bibinfo {author} {\bibfnamefont {S.}~\bibnamefont {Cao}}, \bibinfo {author} {\bibfnamefont {Y.}~\bibnamefont {Zhu}}, \bibinfo {author} {\bibfnamefont {H.}~\bibnamefont {Dong}}, \bibinfo {author} {\bibfnamefont {Y.}~\bibnamefont {Wang}},\ and\ \bibinfo {author} {\bibfnamefont {P.}~\bibnamefont {Qian}},\ }\bibfield  {title} {\bibinfo {title} {Thermal transports of 2d phosphorous carbides by machine learning molecular dynamics simulations},\ }\href@noop {} {\bibfield  {journal} {\bibinfo  {journal} {International Journal of Heat and Mass Transfer}\ }\textbf {\bibinfo {volume} {224}},\ \bibinfo {pages} {125359} (\bibinfo {year} {2024})}\BibitemShut {NoStop}%
\bibitem [{\citenamefont {Fan}\ \emph {et~al.}(2019)\citenamefont {Fan}, \citenamefont {Dong}, \citenamefont {Harju},\ and\ \citenamefont {Ala-Nissila}}]{fan2019homogeneous}%
  \BibitemOpen
  \bibfield  {author} {\bibinfo {author} {\bibfnamefont {Z.}~\bibnamefont {Fan}}, \bibinfo {author} {\bibfnamefont {H.}~\bibnamefont {Dong}}, \bibinfo {author} {\bibfnamefont {A.}~\bibnamefont {Harju}},\ and\ \bibinfo {author} {\bibfnamefont {T.}~\bibnamefont {Ala-Nissila}},\ }\bibfield  {title} {\bibinfo {title} {Homogeneous nonequilibrium molecular dynamics method for heat transport and spectral decomposition with many-body potentials},\ }\href@noop {} {\bibfield  {journal} {\bibinfo  {journal} {Physical Review B}\ }\textbf {\bibinfo {volume} {99}},\ \bibinfo {pages} {064308} (\bibinfo {year} {2019})}\BibitemShut {NoStop}%
\bibitem [{\citenamefont {Evans}(1982)}]{evans1982homogeneous}%
  \BibitemOpen
  \bibfield  {author} {\bibinfo {author} {\bibfnamefont {D.~J.}\ \bibnamefont {Evans}},\ }\bibfield  {title} {\bibinfo {title} {Homogeneous {NEMD} algorithm for thermal conductivity—application of non-canonical linear response theory},\ }\href@noop {} {\bibfield  {journal} {\bibinfo  {journal} {Physics Letters A}\ }\textbf {\bibinfo {volume} {91}},\ \bibinfo {pages} {457} (\bibinfo {year} {1982})}\BibitemShut {NoStop}%
\bibitem [{\citenamefont {Kuhs}\ \emph {et~al.}(1979)\citenamefont {Kuhs}, \citenamefont {Nitsche},\ and\ \citenamefont {Scheunemann}}]{kuhs1979argyrodites}%
  \BibitemOpen
  \bibfield  {author} {\bibinfo {author} {\bibfnamefont {W.}~\bibnamefont {Kuhs}}, \bibinfo {author} {\bibfnamefont {R.}~\bibnamefont {Nitsche}},\ and\ \bibinfo {author} {\bibfnamefont {K.}~\bibnamefont {Scheunemann}},\ }\bibfield  {title} {\bibinfo {title} {The argyrodites—a new family of tetrahedrally close-packed structures},\ }\href@noop {} {\bibfield  {journal} {\bibinfo  {journal} {Materials Research Bulletin}\ }\textbf {\bibinfo {volume} {14}},\ \bibinfo {pages} {241} (\bibinfo {year} {1979})}\BibitemShut {NoStop}%
\bibitem [{\citenamefont {Studenyak}\ \emph {et~al.}(2017)\citenamefont {Studenyak}, \citenamefont {Izai}, \citenamefont {Pogodin}, \citenamefont {Kokhan}, \citenamefont {Sidey}, \citenamefont {Sabov}, \citenamefont {Ke{\v{z}}ionis}, \citenamefont {{\v{S}}alkus},\ and\ \citenamefont {Banys}}]{studenyak2017structural}%
  \BibitemOpen
  \bibfield  {author} {\bibinfo {author} {\bibfnamefont {I.~P.}\ \bibnamefont {Studenyak}}, \bibinfo {author} {\bibfnamefont {V.~Y.}\ \bibnamefont {Izai}}, \bibinfo {author} {\bibfnamefont {A.~I.}\ \bibnamefont {Pogodin}}, \bibinfo {author} {\bibfnamefont {O.~P.}\ \bibnamefont {Kokhan}}, \bibinfo {author} {\bibfnamefont {V.~I.}\ \bibnamefont {Sidey}}, \bibinfo {author} {\bibfnamefont {M.~Y.}\ \bibnamefont {Sabov}}, \bibinfo {author} {\bibfnamefont {A.}~\bibnamefont {Ke{\v{z}}ionis}}, \bibinfo {author} {\bibfnamefont {T.}~\bibnamefont {{\v{S}}alkus}},\ and\ \bibinfo {author} {\bibfnamefont {J.}~\bibnamefont {Banys}},\ }\bibfield  {title} {\bibinfo {title} {Structural and electrical properties of argyrodite-type cu7ps6 crystals},\ }\href@noop {} {\bibfield  {journal} {\bibinfo  {journal} {Lithuanian Journal of Physics}\ }\textbf {\bibinfo {volume} {57}} (\bibinfo {year} {2017})}\BibitemShut {NoStop}%
\bibitem [{\citenamefont {Beeken}\ \emph {et~al.}(2008)\citenamefont {Beeken}, \citenamefont {Driessen}, \citenamefont {Hinaus},\ and\ \citenamefont {Pawlisch}}]{beeken2008electrical}%
  \BibitemOpen
  \bibfield  {author} {\bibinfo {author} {\bibfnamefont {R.}~\bibnamefont {Beeken}}, \bibinfo {author} {\bibfnamefont {C.}~\bibnamefont {Driessen}}, \bibinfo {author} {\bibfnamefont {B.}~\bibnamefont {Hinaus}},\ and\ \bibinfo {author} {\bibfnamefont {D.}~\bibnamefont {Pawlisch}},\ }\bibfield  {title} {\bibinfo {title} {Electrical conductivity of ag7pse6 and cu7pse6},\ }\href@noop {} {\bibfield  {journal} {\bibinfo  {journal} {Solid State Ionics}\ }\textbf {\bibinfo {volume} {179}},\ \bibinfo {pages} {1058} (\bibinfo {year} {2008})}\BibitemShut {NoStop}%
\bibitem [{\citenamefont {Senftle}\ \emph {et~al.}(2016)\citenamefont {Senftle}, \citenamefont {Hong}, \citenamefont {Islam}, \citenamefont {Kylasa}, \citenamefont {Zheng}, \citenamefont {Shin} \emph {et~al.}}]{senftle2016reaxff}%
  \BibitemOpen
  \bibfield  {author} {\bibinfo {author} {\bibfnamefont {T.~P.}\ \bibnamefont {Senftle}}, \bibinfo {author} {\bibfnamefont {S.}~\bibnamefont {Hong}}, \bibinfo {author} {\bibfnamefont {M.~M.}\ \bibnamefont {Islam}}, \bibinfo {author} {\bibfnamefont {S.~B.}\ \bibnamefont {Kylasa}}, \bibinfo {author} {\bibfnamefont {Y.}~\bibnamefont {Zheng}}, \bibinfo {author} {\bibfnamefont {Y.~K.}\ \bibnamefont {Shin}}, \emph {et~al.},\ }\bibfield  {title} {\bibinfo {title} {The reaxff reactive force-field: development, applications and future directions},\ }\href@noop {} {\bibfield  {journal} {\bibinfo  {journal} {npj Computational Materials}\ }\textbf {\bibinfo {volume} {2}},\ \bibinfo {pages} {1} (\bibinfo {year} {2016})}\BibitemShut {NoStop}%
\bibitem [{\citenamefont {Almyras}\ \emph {et~al.}(2019)\citenamefont {Almyras}, \citenamefont {Sangiovanni},\ and\ \citenamefont {Sarakinos}}]{almyras2019semi}%
  \BibitemOpen
  \bibfield  {author} {\bibinfo {author} {\bibfnamefont {G.}~\bibnamefont {Almyras}}, \bibinfo {author} {\bibfnamefont {D.~G.}\ \bibnamefont {Sangiovanni}},\ and\ \bibinfo {author} {\bibfnamefont {K.}~\bibnamefont {Sarakinos}},\ }\bibfield  {title} {\bibinfo {title} {Semi-empirical force-field model for the ti1--xalxn(0$\leq$x$\leq$1) system},\ }\href@noop {} {\bibfield  {journal} {\bibinfo  {journal} {Materials}\ }\textbf {\bibinfo {volume} {12}},\ \bibinfo {pages} {215} (\bibinfo {year} {2019})}\BibitemShut {NoStop}%
\bibitem [{\citenamefont {Butler}\ \emph {et~al.}(2018)\citenamefont {Butler}, \citenamefont {Davies}, \citenamefont {Cartwright}, \citenamefont {Isayev},\ and\ \citenamefont {Walsh}}]{butler2018machine}%
  \BibitemOpen
  \bibfield  {author} {\bibinfo {author} {\bibfnamefont {K.~T.}\ \bibnamefont {Butler}}, \bibinfo {author} {\bibfnamefont {D.~W.}\ \bibnamefont {Davies}}, \bibinfo {author} {\bibfnamefont {H.}~\bibnamefont {Cartwright}}, \bibinfo {author} {\bibfnamefont {O.}~\bibnamefont {Isayev}},\ and\ \bibinfo {author} {\bibfnamefont {A.}~\bibnamefont {Walsh}},\ }\bibfield  {title} {\bibinfo {title} {Machine learning for molecular and materials science},\ }\href@noop {} {\bibfield  {journal} {\bibinfo  {journal} {Nature}\ }\textbf {\bibinfo {volume} {559}},\ \bibinfo {pages} {547} (\bibinfo {year} {2018})}\BibitemShut {NoStop}%
\bibitem [{\citenamefont {Pun}\ \emph {et~al.}(2019)\citenamefont {Pun}, \citenamefont {Batra}, \citenamefont {Ramprasad},\ and\ \citenamefont {Mishin}}]{pun2019physically}%
  \BibitemOpen
  \bibfield  {author} {\bibinfo {author} {\bibfnamefont {G.~P.}\ \bibnamefont {Pun}}, \bibinfo {author} {\bibfnamefont {R.}~\bibnamefont {Batra}}, \bibinfo {author} {\bibfnamefont {R.}~\bibnamefont {Ramprasad}},\ and\ \bibinfo {author} {\bibfnamefont {Y.}~\bibnamefont {Mishin}},\ }\bibfield  {title} {\bibinfo {title} {Physically informed artificial neural networks for atomistic modeling of materials},\ }\href@noop {} {\bibfield  {journal} {\bibinfo  {journal} {Nature communications}\ }\textbf {\bibinfo {volume} {10}},\ \bibinfo {pages} {1} (\bibinfo {year} {2019})}\BibitemShut {NoStop}%
\bibitem [{\citenamefont {Qian}\ \emph {et~al.}(2019)\citenamefont {Qian}, \citenamefont {Peng}, \citenamefont {Li}, \citenamefont {Wei},\ and\ \citenamefont {Yang}}]{qian2019thermal}%
  \BibitemOpen
  \bibfield  {author} {\bibinfo {author} {\bibfnamefont {X.}~\bibnamefont {Qian}}, \bibinfo {author} {\bibfnamefont {S.}~\bibnamefont {Peng}}, \bibinfo {author} {\bibfnamefont {X.}~\bibnamefont {Li}}, \bibinfo {author} {\bibfnamefont {Y.}~\bibnamefont {Wei}},\ and\ \bibinfo {author} {\bibfnamefont {R.}~\bibnamefont {Yang}},\ }\bibfield  {title} {\bibinfo {title} {Thermal conductivity modeling using machine learning potentials: application to crystalline and amorphous silicon},\ }\href@noop {} {\bibfield  {journal} {\bibinfo  {journal} {Materials Today Physics}\ }\textbf {\bibinfo {volume} {10}},\ \bibinfo {pages} {100140} (\bibinfo {year} {2019})}\BibitemShut {NoStop}%
\bibitem [{\citenamefont {Fan}\ \emph {et~al.}(2021)\citenamefont {Fan}, \citenamefont {Zeng}, \citenamefont {Zhang}, \citenamefont {Wang}, \citenamefont {Song}, \citenamefont {Dong}, \citenamefont {Chen},\ and\ \citenamefont {Ala-Nissila}}]{fan2021neuroevolution}%
  \BibitemOpen
  \bibfield  {author} {\bibinfo {author} {\bibfnamefont {Z.}~\bibnamefont {Fan}}, \bibinfo {author} {\bibfnamefont {Z.}~\bibnamefont {Zeng}}, \bibinfo {author} {\bibfnamefont {C.}~\bibnamefont {Zhang}}, \bibinfo {author} {\bibfnamefont {Y.}~\bibnamefont {Wang}}, \bibinfo {author} {\bibfnamefont {K.}~\bibnamefont {Song}}, \bibinfo {author} {\bibfnamefont {H.}~\bibnamefont {Dong}}, \bibinfo {author} {\bibfnamefont {Y.}~\bibnamefont {Chen}},\ and\ \bibinfo {author} {\bibfnamefont {T.}~\bibnamefont {Ala-Nissila}},\ }\bibfield  {title} {\bibinfo {title} {Neuroevolution machine learning potentials: Combining high accuracy and low cost in atomistic simulations and application to heat transport},\ }\href@noop {} {\bibfield  {journal} {\bibinfo  {journal} {Physical Review B}\ }\textbf {\bibinfo {volume} {104}},\ \bibinfo {pages} {104309} (\bibinfo {year} {2021})}\BibitemShut {NoStop}%
\bibitem [{\citenamefont {Plimpton}(1995)}]{plimpton1995fast}%
  \BibitemOpen
  \bibfield  {author} {\bibinfo {author} {\bibfnamefont {S.}~\bibnamefont {Plimpton}},\ }\bibfield  {title} {\bibinfo {title} {Fast parallel algorithms for short-range molecular dynamics},\ }\href@noop {} {\bibfield  {journal} {\bibinfo  {journal} {Journal of computational physics}\ }\textbf {\bibinfo {volume} {117}},\ \bibinfo {pages} {1} (\bibinfo {year} {1995})}\BibitemShut {NoStop}%
\bibitem [{\citenamefont {Kresse}\ and\ \citenamefont {Furthm{\"u}ller}(1996)}]{kresse1996efficient}%
  \BibitemOpen
  \bibfield  {author} {\bibinfo {author} {\bibfnamefont {G.}~\bibnamefont {Kresse}}\ and\ \bibinfo {author} {\bibfnamefont {J.}~\bibnamefont {Furthm{\"u}ller}},\ }\bibfield  {title} {\bibinfo {title} {Efficient iterative schemes for $ab$ $initio$ total-energy calculations using a plane-wave basis set},\ }\href@noop {} {\bibfield  {journal} {\bibinfo  {journal} {Physical Review B}\ }\textbf {\bibinfo {volume} {54}},\ \bibinfo {pages} {11169} (\bibinfo {year} {1996})}\BibitemShut {NoStop}%
\bibitem [{\citenamefont {Bl{\"o}chl}(1994)}]{blochl1994projector}%
  \BibitemOpen
  \bibfield  {author} {\bibinfo {author} {\bibfnamefont {P.~E.}\ \bibnamefont {Bl{\"o}chl}},\ }\bibfield  {title} {\bibinfo {title} {Projector augmented-wave method},\ }\href@noop {} {\bibfield  {journal} {\bibinfo  {journal} {Physical Review B}\ }\textbf {\bibinfo {volume} {50}},\ \bibinfo {pages} {17953} (\bibinfo {year} {1994})}\BibitemShut {NoStop}%
\bibitem [{\citenamefont {Perdew}\ \emph {et~al.}(2008)\citenamefont {Perdew}, \citenamefont {Ruzsinszky}, \citenamefont {Csonka}, \citenamefont {Vydrov}, \citenamefont {Scuseria}, \citenamefont {Constantin}, \citenamefont {Zhou},\ and\ \citenamefont {Burke}}]{perdew2008restoring}%
  \BibitemOpen
  \bibfield  {author} {\bibinfo {author} {\bibfnamefont {J.~P.}\ \bibnamefont {Perdew}}, \bibinfo {author} {\bibfnamefont {A.}~\bibnamefont {Ruzsinszky}}, \bibinfo {author} {\bibfnamefont {G.~I.}\ \bibnamefont {Csonka}}, \bibinfo {author} {\bibfnamefont {O.~A.}\ \bibnamefont {Vydrov}}, \bibinfo {author} {\bibfnamefont {G.~E.}\ \bibnamefont {Scuseria}}, \bibinfo {author} {\bibfnamefont {L.~A.}\ \bibnamefont {Constantin}}, \bibinfo {author} {\bibfnamefont {X.}~\bibnamefont {Zhou}},\ and\ \bibinfo {author} {\bibfnamefont {K.}~\bibnamefont {Burke}},\ }\bibfield  {title} {\bibinfo {title} {Restoring the density-gradient expansion for exchange in solids and surfaces},\ }\href@noop {} {\bibfield  {journal} {\bibinfo  {journal} {Physical Review Letters}\ }\textbf {\bibinfo {volume} {100}},\ \bibinfo {pages} {136406} (\bibinfo {year} {2008})}\BibitemShut {NoStop}%
\bibitem [{\citenamefont {Gubaev}\ \emph {et~al.}(2019)\citenamefont {Gubaev}, \citenamefont {Podryabinkin}, \citenamefont {Hart},\ and\ \citenamefont {Shapeev}}]{gubaev2019accelerating}%
  \BibitemOpen
  \bibfield  {author} {\bibinfo {author} {\bibfnamefont {K.}~\bibnamefont {Gubaev}}, \bibinfo {author} {\bibfnamefont {E.~V.}\ \bibnamefont {Podryabinkin}}, \bibinfo {author} {\bibfnamefont {G.~L.}\ \bibnamefont {Hart}},\ and\ \bibinfo {author} {\bibfnamefont {A.~V.}\ \bibnamefont {Shapeev}},\ }\bibfield  {title} {\bibinfo {title} {Accelerating high-throughput searches for new alloys with active learning of interatomic potentials},\ }\href@noop {} {\bibfield  {journal} {\bibinfo  {journal} {Computational Materials Science}\ }\textbf {\bibinfo {volume} {156}},\ \bibinfo {pages} {148} (\bibinfo {year} {2019})}\BibitemShut {NoStop}%
\bibitem [{\citenamefont {Dongre}\ \emph {et~al.}(2017)\citenamefont {Dongre}, \citenamefont {Wang},\ and\ \citenamefont {Madsen}}]{dongre2017comparison}%
  \BibitemOpen
  \bibfield  {author} {\bibinfo {author} {\bibfnamefont {B.}~\bibnamefont {Dongre}}, \bibinfo {author} {\bibfnamefont {T.}~\bibnamefont {Wang}},\ and\ \bibinfo {author} {\bibfnamefont {G.~K.}\ \bibnamefont {Madsen}},\ }\bibfield  {title} {\bibinfo {title} {Comparison of the green--kubo and homogeneous non-equilibrium molecular dynamics methods for calculating thermal conductivity},\ }\href@noop {} {\bibfield  {journal} {\bibinfo  {journal} {Modelling and Simulation in Materials Science and Engineering}\ }\textbf {\bibinfo {volume} {25}},\ \bibinfo {pages} {054001} (\bibinfo {year} {2017})}\BibitemShut {NoStop}%
\bibitem [{\citenamefont {Ouyang}\ \emph {et~al.}(2022{\natexlab{b}})\citenamefont {Ouyang}, \citenamefont {Wang},\ and\ \citenamefont {Chen}}]{ouyang2022temperature}%
  \BibitemOpen
  \bibfield  {author} {\bibinfo {author} {\bibfnamefont {N.}~\bibnamefont {Ouyang}}, \bibinfo {author} {\bibfnamefont {C.}~\bibnamefont {Wang}},\ and\ \bibinfo {author} {\bibfnamefont {Y.}~\bibnamefont {Chen}},\ }\bibfield  {title} {\bibinfo {title} {Temperature-and pressure-dependent phonon transport properties of {SnS} across phase transition from machine-learning interatomic potential},\ }\href@noop {} {\bibfield  {journal} {\bibinfo  {journal} {International Journal of Heat and Mass Transfer}\ }\textbf {\bibinfo {volume} {192}},\ \bibinfo {pages} {122859} (\bibinfo {year} {2022}{\natexlab{b}})}\BibitemShut {NoStop}%
\end{thebibliography}%

\end{document}